\def\ap3m{AP$^3$M}

\def\Msun{${\rm M_{\odot}}$~}

\def\ea{et~al.~}                            

\def\lesssim{\mathrel{\hbox{\rlap{\hbox{\lower4pt\hbox{$\sim$}}}\hbox{$<$}}}}
\def\grtsim{\mathrel{\hbox{\rlap{\hbox{\lower4pt\hbox{$\sim$}}}\hbox{$>$}}}}

\newcommand{\AAA}[3]    {\mbox{#3, A\&A,~#1,~#2}}

\newcommand{\ApJ}[3]    {\mbox{#3, ApJ,~#1,~#2}}
\newcommand{\ApJS}[3]   {\mbox{#3, ApJ~Suppl.,~#1,~#2}}

\newcommand{\AJ}[3]     {\mbox{#3, Astron.~J.,~#1,~#2}}
\newcommand{\MNRAS}[3]  {\mbox{#3, MNRAS,~#1,~#2}}
\newcommand{\Nature}[3] {\mbox{#3, Nature,~#1,~#2}}
\newcommand{\NewA}[3]   {\mbox{#3, NewA,~#1,~#2}}

\newcommand{\astroph}[1]{\mbox{preprint~(astro-ph/#1)}}

\newcommand{\bh}	{\bullet}
\newcommand{\BH}	{\mbox{\boldmath$ \bullet$}}

\documentstyle[epsfig]{mn}

\begin{document}

\title{Massive black hole remnants of the first stars in galactic haloes}

\author[Islam R., Taylor J. \& Silk J.]
       {Ranty R. Islam, James E. Taylor and Joseph Silk\\        
       {Astrophysics, Denys Wilkinson Building, Keble Road, Oxford, OX1 3RH, UK}}

\date{Received ...; accepted ...}

\maketitle

\begin{abstract}

We investigate the possibility that present-day galactic haloes contain 
a population of massive black holes (MBHs) that form by
hierarchical merging of the black hole remnants of the first stars.
Some of the MBHs may be large enough or close enough to the
centre of the galactic host that they merge within a Hubble time. We
estimate to what extent this process could contribute to the mass 
of the super-massive black holes (SMBHs) observed in galactic
centres today. Many MBHs will not reach the centre of the main halo,
however, but continue to orbit within satellite subhaloes.
Using a semi-analytical approach that explicitly accounts for dynamical
friction, tidal disruption and encounters with the galactic disk, we
follow the dynamics of the satellites and their MBHs and determine the 
abundance and distribution of MBHs in present-day haloes of various
masses.
Considering two different accretion scenarios we also compute the bolometric
luminosity function for the MBHs.
\end{abstract}

\begin{keywords}
cosmology: theory -- 
galaxies: formation -- galaxies: nuclei
\end{keywords}

\section{Introduction}
The presence of super-massive black holes (SMBHs) at the centres of most
galaxies appears by now firmly established. SMBHs have estimated
masses in the range $10^6 - 10^9$ ~\Msun and various correlations
have been observed between the mass of 
SMBHs and properties of the galactic bulge hosting them.
The first of these to be established were correlations between the
mass of the SMBH, $M_{smbh}$ and the mass or luminosity of the
galactic bulge,  $M_{bulge}$ and $L_{bulge}$ 
respectively \cite{magorrian98,kormendy00,laor01}.
More recently, a tighter correlation was found between $M_{smbh}$ and
the bulge velocity dispersion, $\sigma_{bulge}$ at some fiducial distance from the
centre \cite{gebhardt00,merritt01}.
An equally tight correlation has also been determined between 
$M_{smbh}$ and the bulge's light profile, as parameterised by a shape
index, $n$ \cite{graham01}.

Since these correlations extend well beyond the direct dynamical
influence  of the SMBH it seems likely that there is a close link
between the formation of SMBHs and the formation of their host galaxy.
A recent analysis finds that the masses of SMBHs appear to be
correlated with the host circular velocity even beyond the optical
radius \cite{ferrarese02}. If this is confirmed, it indicates that the SMBHs are linked
to properties of the host dark matter halo. This would be the
strongest hint yet that there must be a hierarchical merging component to
the growth of SMBHs, since the properties of halos are primarily
determined in the context of their hierarchical build up.

Most models put
forward to account for the correlations assume a close link between
galaxy and SMBH formation as a starting point, although they subsequently
proceed along either or both of two routes to explain how the SMBHs
grow in mass. 
One is to consider that the SMBH mass increases mainly by the merging
of smaller precursors. 
This requires SMBH precursors to have been present in galaxies from very
early on \cite{madau01,menou01,schneider02} 
It might allow the observed correlations to be set up over a long period 
of time with a potentially large number of mergers through the
dynamical interactions between the merging galaxies and SMBH
precursors. However, BH merging by itself might ultimately be highly inefficient
especially for low mass BH binary systems, for which it would be extremely
difficult to progress from a mutually bound configuration to
the stage where emission of gravitational radiation draws the binary
constituents to final coalescence.     
 
Another mechanism considered is growth mainly by gas accretion within the 
host bulge. In this case a strong non-gravitational interaction between the
growing SMBH and the bulge has to be invoked. An example of this is
the radiative feedback of an accreting SMBH that changes the gas dynamics
in the bulge so as to effectively control its own gas supply and establish a
relation between $M_{smbh}$ and $\sigma_{bulge}$ \cite{silk98}.
A similar route is followed by  models that tie $M_{smbh}$ to the amount and properties of gas
in the bulge \cite{adams01}. The latter itself may depend on the
previous merging of the galaxy with others and so provides a way of
combining SMBH mass growth through both mergers and accretion
\cite{haehnelt00}.

As an example of the merger-only scenario it has been shown that the
merging of the massive black hole (MBH) remnants of the first stars in
the Universe could account for the inferred overall abundance of SMBHs
today \cite{schneider02}.

However, gas accretion during the optically bright QSO phase may be
able to account for most of the present day SMBH mass density
\cite{yu02b}, although this process alone would probably not allow
ordinary stellar mass BHs to become as large as the most massive SMBHs
observed today ($M_{smbh} \grtsim 10^9$ \Msun) (see e.g. Richstone
~\ea 1998). Even if stellar mass BHs were accreting at the
Eddington limit, there would not be enough time for the required mass
increase to occur. The presence of massive BH seeds at prior to
the QSO phase and/or subsequent merging of MBHs therefore appears to be necessary. 
 
In this paper we explore this idea further to determine an upper limit
on the mass to which SMBHs can grow through mergers of
lower mass precursors and more importantly what the implications are for the presence of a remnant 
population of lower mass MBHs in the galactic halo. In doing so we
assume efficient merging between MBHs, but we also consider the effect
of relaxing this assumption.
As the `seeds' in the merging hierarchy, we consider massive
black holes (MBHs) of some mass $M_{seed}$ that are remnants of the
first stars in the Universe, forming within high-$\sigma$ density
peaks at redshifts of $z \sim 24$. We use
Monte Carlo merger trees to describe the merging of haloes and then
follow the dynamical evolution of merged/accreted satellite haloes and
their central MBHs within larger hosts, explicitly accounting for dynamical
friction, tidal stripping and disk encounters. 
A key prediction is that $\sim 10^3$ MBHs in the mass range $1 -
1000 \times M_{seed}$  should be present within the galactic halo today as a
result of this process. 

We start by describing the origin of seed MBHs in section
\ref{sec:mbhorigin}. In section \ref{sec:mbhmerge}, we explain how the subsequent
merging of their haloes could lead to a build-up of a population
of MBHs in present-day galactic haloes, as well as contribute to the mass of a central SMBH. 
Ways of detecting the population of halo
MBHs, particularly via their X-ray emission, are described in section
\ref{sec:mbhdetect}. We conclude with a
summary of our findings in section \ref{sec:summary}. 
 
\section{Primordial Star Formation and Massive Black Holes} \label{sec:mbhorigin}
A number of recent semi-analytical \cite{hutchings02,fuller00,tegmark97} and
numerical investigations \cite{bromm02,abel00} suggests that the
first stars in the Universe were likely created inside molecular
clouds that fragmented out of the first baryonic cores 
inside dark matter halos at very high redshifts.
For common $\Lambda$CDM cosmologies in particular these objects are
found to have a mass $M_{min} \sim 3 \times 10^5 h^{-1}$ \Msun and to
have collapsed at redshift $z
\sim 24$. In linear collapse theory this corresponds to collapse from
$3 \sigma$ peaks in the initial matter density field. This is because the mass
contained in overdensities corresponding to $3 \sigma$ peaks at this
redshift is just higher than both the cosmological Jeans mass and the 
cooling mass \footnote{At or above the cooling mass the corresponding
virial temperature, to which the baryons are heated, is high enough for
cooling to proceed on a time scale that is smaller than the gravitational infall
time scale. The latter is the condition for fragmentation to occur.}
Cooling nevertheless proceeds much more slowly than at present; as
stars have yet to form, metals that could facilitate more efficient 
cooling are essentially not present. 
This implies that even though fragmentation occurs, fragments will be
much larger than in a corresponding situation today. 
Seed masses within these 
fragments can in principle accrete large amounts of matter from the
cloud without further fragmentation occurring, which could eventually
lead to the formation of a proto-star.
\cite{bromm02,machacek01,omukai01}. Only radiation pressure from 
the proto-star on the infalling layers of material could halt
accretion and so limit the mass of the star. However, in the absence of dust the
infalling matter has too low an opacity for radiation pressure to be
significant \cite{ripamonti01}. In these stars the role of winds that 
could lead to significant mass loss in population I stars, is also negligible.
As a result this will likely lead to the creation of very massive
stars, potentially as heavy as $10^4$ \Msun. These are also referred
to as population III stars. 

As yet nothing definite is known about the initial mass function (IMF) 
of these stars. However, their large mass will see 
many of them ending up as black holes of essentially the same mass -
gravity is so strong that not even ejecta of a final supernova can
escape \cite{heger01}. 

Here we assume that in each dark matter halo forming at $z = 24$ with
a mass larger than $3 \times 10^5 h^{-1}$ \Msun, one MBH of mass
260 \Msun forms as the end result of any primordial star formation
occurring inside the halo.
These MBHs then form the seeds for the subsequent merging process.

\subsection{MBH mass density}
Before considering the merger of MBHs after their formation
and their subsequent dynamical evolution within merged host haloes it
is instructive to look at the global mass densities contained in MBHs.
If one MBH of mass $260$ \Msun forms in each halo corresponding to
$3 \sigma$ peaks or higher in the initial matter density field at
redshift $z \sim 24$ the mass density in MBH is (see also Madau \&
Rees 2001 for the case of a SCDM cosmology)
\begin{equation}
	\rho_{\bh} \le \frac{0.0027 \Omega_0 \rho_{crit} m_{\bh}}{10^5
{\rm M}_{\odot} {\rm Mpc^{-3}}} \approx 2.9 \times 10^5 {\rm
M}_{\odot} {\rm Mpc}^{-3}
\end{equation}
based on a $\Lambda$CDM cosmology with  $\Omega_0 = 0.3,
\Omega_{\Lambda} = 0.7, h = 0.7$. Our subsequent analysis is also based
on this cosmology.
For the actual local mass density contained in SMBHs Merritt \&
Ferrarese (2001) obtain $\rho_{\bh} \approx 5 \times 10^5 {\rm
M}_{\odot} {\rm Mpc}^{-3}$.

This means that in order for SMBH to have grown primarily by mergers of
lower mass MBHs, the mass of the initial seed MBHs need to be just
about twice as massive, i.e. around 500 \Msun
and that most of them end up in SMBH today. 
For less massive seed MBHs, growth through  gas accretion will have to
play an important role in achieving the present-day SMBH mass density. 
Especially the assumption that all MBHs merge to form
the SMBHs, however, is inappropriate and the dynamics of individual
MBHs needs to be examined in more detail as we describe in the next section. 


\section{Hierarchical Merging of Primordial Black Holes} 
\label{sec:mbhmerge}
\subsection{Modelling halo growth}
While the basic properties of the seed MBHs are determined by the 
physics of the first baryonic objects, as outlined above, the extent
to which they merge to form the present-day SMBH depends on their 
subsequent dynamical evolution after their respective host haloes 
have merged. To track this evolution we use a semi-analytical code 
(Taylor \& Babul in preparation; see also Taylor 2001 and Taylor \&
Babul 2001) 
that
combines a Monte-Carlo algorithm to generate halo merger trees with 
analytical descriptions for the main dynamical processes -- 
dynamical friction, tidal stripping, and tidal heating -- 
that determine the evolution of merged remnants within a galaxy halo.

Starting with a halo of a specific mass at the present-day, we trace
the merger history of the system back to a redshift of 30, using
the algorithm of Somerville \& Kollatt (1999). Computational 
considerations limit the mass resolution of the tree to 
$\simeq\,3\times\,10^{-5}$ of the total mass; below this limit 
we do not trace the merger history fully. For the more massive
halos, this resolution limit is larger than $M_{min}$ and many 
of the branchings of the merger tree drop below the mass resolution 
limit before they reach $z = 24$, so that we cannot always track the 
formation of individual black holes. To overcome this problem, if 
systems over $M_{min}$ appear in the merger tree after primordial 
black holes have started forming at $z = 24$, we determine how
likely they are to contain one or more primordial black holes, 
based on the frequency of 3 $\sigma$ peaks, and populate them accordingly. 
In the most massive trees, haloes at the resolution limit are likely
to contain several primordial black holes. In this case, we assume the 
black holes have merged to form a single object, in keeping with
the assumption of efficient merging discussed below.

Within the merger trees, we then follow the dynamical evolution of
black holes forward in time to the present-day, using the analytic model
of satellite dynamics developed in Taylor \& Babul (2001). 
Merging subhaloes are placed on realistic orbits at the virial radius of 
the main system, and experience dynamical friction, mass loss and heating
as they move through their orbits. The background potential is modelled
by a smooth Moore profile ($\rho \propto r^{-1.5}(r_s^{-1.5} + 
r^{-1.5})$), which grows in mass according to 
its merger history, and changes in concentration following the relations 
proposed by Eke, Navarro \& Steinmetz (2001). We give this profile a 
constant-density core
of radius $0.1 r_s$, to account for the possible effects of galaxy 
formation in disrupting the dense central cusp. 

Within this potential, the formation of a central 
galaxy with a disk and a spheroidal component is modelled schematically,
by assuming that a third of the gas within the halo cools on the dynamical
time-scale to form a galactic disk, and that major mergers disrupt this 
disk and transform it into a spheroid with some overall efficiency. 
We choose as the disruption criterion that the disk collide with
an infalling satellite of mass equal to or greater than its own, 
and set the efficiency with which disk material is then transferred 
to the spheroid to 0.25. This choice of parameters is required to
limit the formation of spheroids and thus produce a reasonable range 
of morphologies in isolated present-day $10^{12} M_{\odot}$ systems, 
as discussed in Taylor \& Babul (2002). We do not expect the results 
for halo back holes to depend strongly on these parameters, although
they may have some effect on the properties of the central black holes. 
Finally, the evolution of 
haloes in side branches of the merger tree is followed more approximately,
by assuming that higher-order substructure (that is subhaloes within subhaloes)
merges over a few dynamical times, causing its black hole component to 
merge as well, while unmerged substructure percolates down to a
lower level in the tree. We will discuss the details
of this model in forthcoming work (Taylor \& Babul in preparation);
here it serves only as a backdrop for the dynamical calculations of
black hole evolution. 

The semi-analytic code tracks the positions of all the primordial black holes 
that merge with the main system and the amount of residual dark matter 
from their original halo that still surrounds them, if any. 
We classify systems as `naked' if their surrounding 
subhalo has been completely stripped by tidal forces, and `normal'
otherwise. Our orbital calculations cannot follow the evolution of
systems down to arbitrarily small radii within the main potential, so if 
black holes come within
1\% of the virial radius of the centre of the potential (roughly 3 kpc
for a system like the present-day Milk Way), we assume they have `fallen in'
and stop tracking their orbits. Black holes contained in satellites
which disrupt the disk in major mergers are also assumed to fall into
the centre of the potential during its subsequent rearrangement. 
Clearly, this assumes that black hole merging in the centre of the main 
system is completely efficient, so it will produce a conservative upper 
limit on how many black holes merge with the central SMBH. We discuss
the effect of relaxing these assumptions below. 

Using the semi-analytic code, we generate sets of different realisations 
for seed black hole masses of 260 \Msun ~and 1300 \Msun, 
and for final halo masses of $1.6\times10^{10}$, $1.6\times10^{11}$, 
$1.6\times10^{12}$ and $1.6\times10^{13}$ \Msun.

\begin{figure}{}
      \centerline{\resizebox{\hsize}{!}{\includegraphics{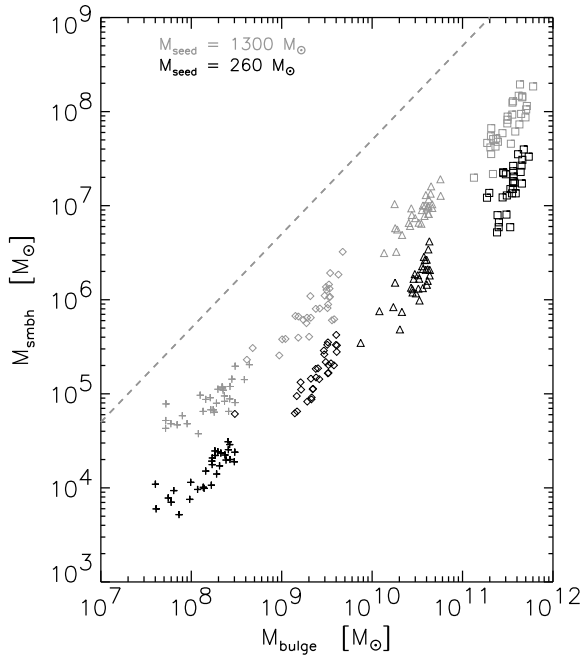}}}
      \caption{Mass of central SMBH versus bulge mass of host
galaxy at z=0. For  bottom left to top right the symbols are for
bulges in haloes with masses $1.6\times 10^{10},1.6\times
10^{11},1.6\times 10^{12},1.6\times 10^{13}$ \Msun. The upper and
lower data sets are for seed MBH masses of $260$ and $1300$ \Msun
~respectively. The observational relation between $M_{bulge}$ and
$M_{smbh}$ is shown by the dashed grey line.}
      \label{fig:MsmbhMbulge}
\end{figure}

\subsection{Central super-massive black holes (SMBH)}
The assumption that MBHs within a kpc or so from the host centre merge 
efficiently can be used to determine an upper limit on the mass of
central SMBHs. Although MBH merging may proceed much less
efficiently, we give examples of a range of processes that can lead to
rapid merging of MBHs in the galactic context. 

If the mass of only the MBHs is considered their orbital decay time
scale in the host can be longer than a Hubble time.
However, MBHs typically remain associated with stars and gas from
their original satellite, which increases their effective mass by a
factor of at least 100 to 1000 and lowers the orbital decay time scale
accordingly, allowing even relatively light MBHs ($M_{\bh} \ge 10^3$
\Msun) to spiral into the host central
region ($\le$ kpc) within a Hubble time \cite{yu02}. This is true even if the
satellite itself may have actually lost most of its mass ($\ge$ 99 percent)
due to tidal stripping inside the host halo and is thus classified as
`naked' in our treatment.
This implies that only at high redshifts could seed mass MBHs have travelled to the host 
centre, since they would have then entered the correspondingly
smaller host halo at smaller distances from the centre.

It seems then that dynamical friction can deliver MBHs to the host central
regions efficiently where they then form binaries with any MBH already 
at the centre. The evolution of a MBH binary system in stellar
background has been studied
extensively \cite{begelman80,quinlan96,milosavljevic01,yu02} and the
`hardening' stage of binary evolution has been
singled out as the `bottle neck' on the way to the final merger
\cite{milosavljevic01,yu02}. With dynamical friction no longer significant
and orbital decay due to gravitational wave emission not yet
important, the only way for the binary MBHs to reduce their orbital
axis is
by interaction with stars in their vicinity, which can take
significantly longer than a Hubble time.

However, the presence of gas may be of crucial
importance in this context \cite{milosavljevic01}. High densities of gas between the binary
MBHs could allow for a much faster evolution and eventual 
merger of the binary. Several scenarios have been suggested for this,
such as a massive gas disk around the binary \cite{gould00} or massive 
gas inflow (see e.g. Begelman, Blandford \& Rees 1980) in the wake of major mergers.
Hydrodynamical simulations of galaxy mergers, for instance,  find that up to 60 per cent of the total gas mass of two
merging Milky Way sized galaxies can end up within a region only a few hundred
parsecs across, which is about half the bulge scale radius
\cite{barnes96,naab01,barnes02}. 
 
Here we assume that during major mergers the gas infall will actually
lead to all MBHs binaries merging. We also neglect the possibility or
triple BH interactions and sling-shot ejections.

Figure \ref{fig:MsmbhMbulge} shows the relation
between the mass of the galactic bulge and the central SMBH if the
latter grows purely through mergers of smaller MBH.
The solid line represents the linear relationship between SMBH and
bulge mass as determined by observations \cite{magorrian98} and here
is only shown to give an upper limit on the allowed masses for the
SMBHs and also the mass of the seed MBHs provided $M_{smbh} \propto
M_{\bh,seed}$
For instance, in the extreme case where the 
seed MBH mass is equal to the total baryonic mass ($\sim
1.3\times10^4$ \Msun) of the $3 \sigma$
haloes within which they first formed, the resulting SMBH masses are
essentially ruled out by the observed relation. 
For both seed masses we are considering,
growth through gas accretion is required to match the observations. 
We also note that a power law best fit between the SMBH and bulge
mass for both seed masses, yields an index $\sim 0.9$, i.e. less than 1, as would be the case for a linear 
relationship. 
This means that for larger bulges a slightly smaller
fraction of the total mass contained in MBHs merges with the SMBH and
therefore a relatively larger amount has to be acquired through gas
accretion to achieve a linear relationship. Larger bulges will have typically formed inside
correspondingly more massive host galaxies/haloes which collapse at
higher redshifts, which implies that more gas must have been available then to be 
accreted by the central SMBH. This trend seems plausible also in the
light of results from star formation and quasar activity at high redshifts.

\subsection{Abundance of MBHs in galactic haloes}
\begin{figure*}
\hbox{\epsfxsize 8.5cm
\epsfbox{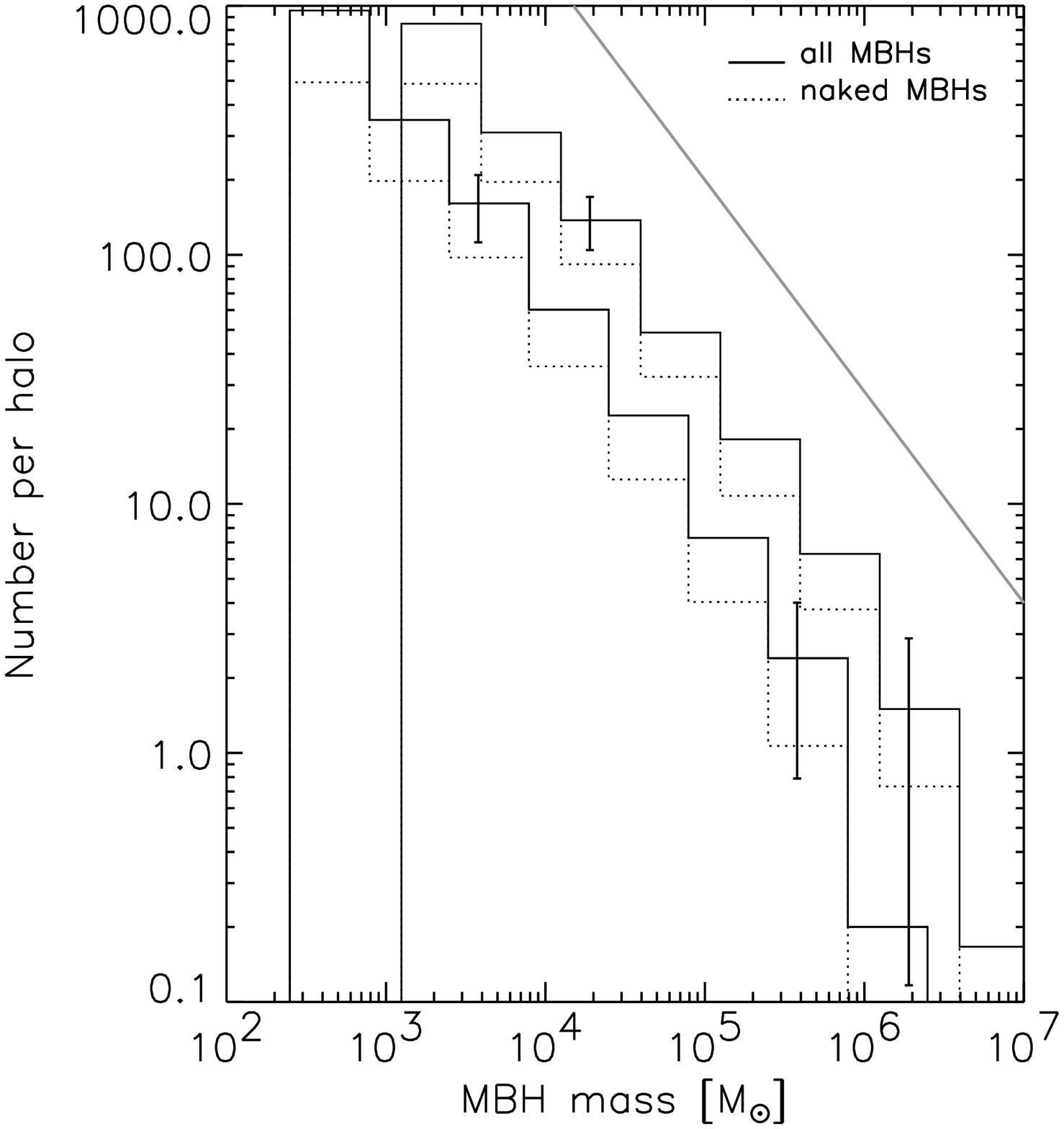}
\hskip 5pt
\epsfxsize 8.5cm
\epsfbox{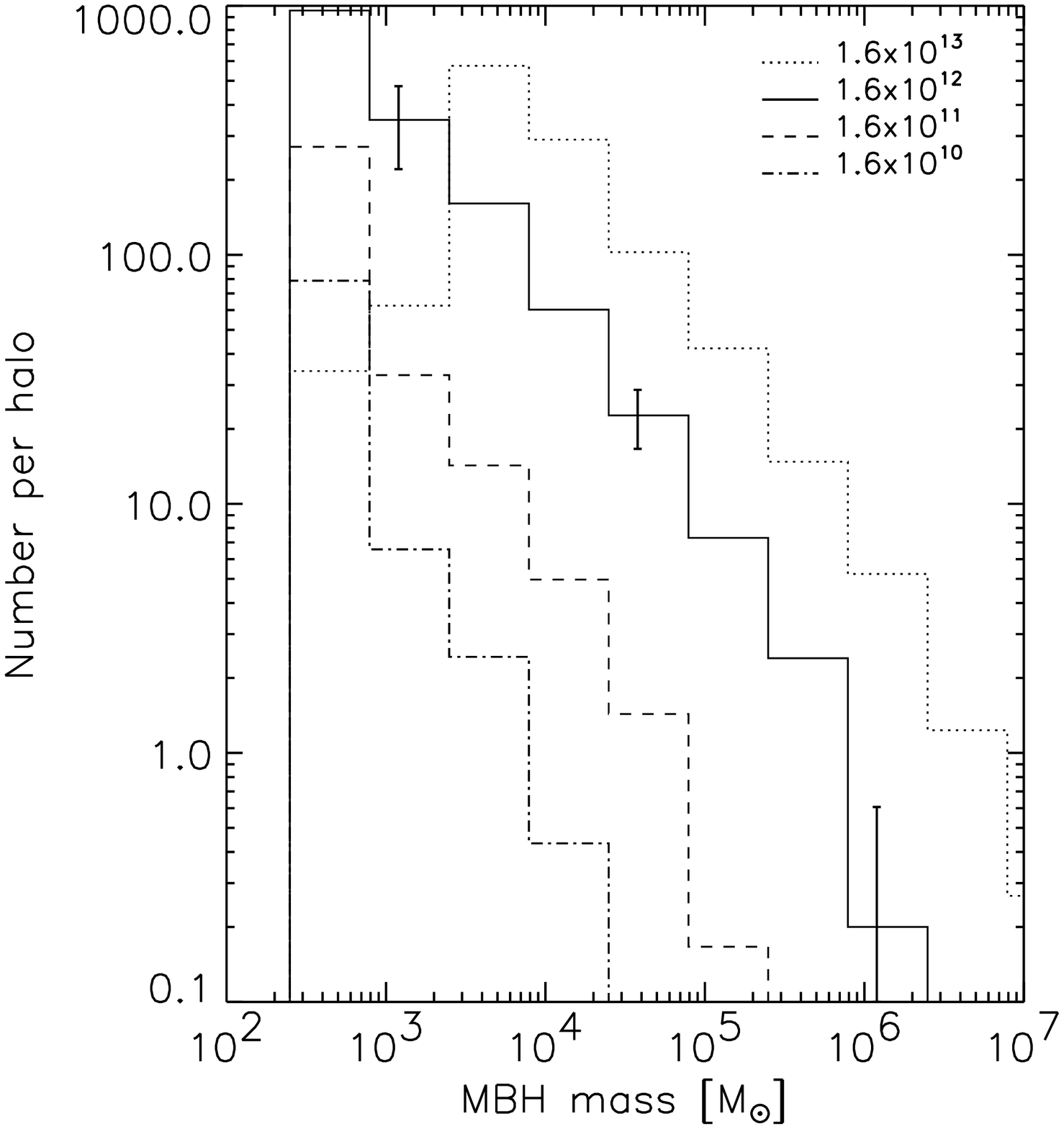}}
\caption{~Mass function of MBHs in the halo
averaged over 30 realisations with error bars corresponding to 1$\sigma$
variance. In the left panel the mass function is shown for the
$1.6\times10^{12}$ \Msun halo for all MBHs as well as only naked
MBHs. Upper and lower data sets are for $M{\bh,seed} = 260$ and $1300$
\Msun respectively. In the right panel the mass functions for
$M{\bh,seed} = 260$ \Msun and all final halo masses are shown.}
\label{fig:abund_mass}
\end{figure*}

In figure \ref{fig:abund_mass} we show the average abundance of  MBHs 
within the virial radius of the primary host halo
and also identify the abundance due to naked MBHs. 

Compared to the mass of the bulge, disk and halo the seed MBH masses
are small and so do not significantly affect the evolution of substructure within the
host. For this reason we find that, except for the high mass end, the
MBH mass functions for the two different MBH seed masses  are essentially
the same but are offset from one another along the ordinate (representing the
actual MBH mass) by a constant factor that is more or less equal to the ratio of the initial 
seed MBH masses. Based on this the solid line in figure
\ref{fig:abund_mass} represents the inferred mass function for a seed MBH with a
mass of $1.3\times10^{4}$ \Msun, that is the case where the entire
primordial cloud collapses into the black hole.
Because the different seed MBH masses only become important at the
high mass end, this scaling just reflects the one-to-one
correspondence between number and mass of $3 \sigma$ haloes and their
seed MBHs.
In the following our analysis will therefore focus on 
the case of 260 \Msun seed mass MBHs unless stated otherwise.

For final halo masses of $1.6\times10^{11},1.6\times10^{12}$ and
$1.6\times10^{13}$ \Msun the number $N$ of remnant MBHs in the halo follows a power law
\begin{equation}
N \sim M_{\bh}^{-0.79 \pm 0.04}
\end{equation}
whereas for the $1.6\times10^{10}$ \Msun halo this is more uncertain,
$N \sim M_{\bh}^{-0.97 \pm 0.1}$. 
The total number of MBHs in the halo is given in table
\ref{tab:global_abund}. For Milky Way sized haloes, for instance, we would expect 
some 1400 to 1500 MBHs to orbit within the galactic halo. We found
that the number of MBHs in the galactic disk out to about two disk
scale radii is less than 0.2 per cent of the total number of MBHs for
all final halo masses. Part of the reason for this low number is that
a lot of the MBHs in the disk are orbiting at small distances of less
than 1 per cent of the host virial radius and are therefore counted as
having fallen to the centre since their dynamics cannot be traced
accurately any more as mentioned above.
Conversely the high mass end implies that apart from the central SMBH
there will be one or two other MBH of about a tenth of its mass orbiting in the halo 

\begin{table}
\caption{Total number of MBHs in halo averaged over thirty 
trees with associated variance.}
\label{tab:global_abund}
\begin{tabular}{llll} \hline
Halo mass & \multicolumn{2}{c}{$M_{\bh,seed}$} \\
 & 260 \Msun & 1300 \Msun \\ \hline \hline
$1.6\times10^{10}$ & 88 $\pm$ 22 & 72 $\pm$ 16 \\
$1.6\times10^{11}$ & 330 $\pm$ 50 & 420 $\pm$ 70 \\
$1.6\times10^{12}$ & 1560 $\pm$ 550 & 1370 $\pm$ 340 \\
$1.6\times10^{13}$ & 1130 $\pm$ 100 & 1430 $\pm$ 310 \\
\hline
\end{tabular}
\end{table}

\begin{table}
\caption{Abundance of MBH in Earth-centred volumes at 8.5 kpc from the 
galactic centre in the Milky-Way-sized halo ($1.6\times10^{12}$ \Msun). Given are the
average over thirty trees with their respective variance.}
\label{tab:local_abund}
\begin{tabular}{llll}  \hline
 $M_{\bh,seed}$  & \multicolumn{3}{c}{Distance from Earth $\Delta r$ [kpc]}\\
 & 2.0  & 2.5  & 3.0  \\ \hline \hline
260 \Msun & 1.12 $\pm$ 0.6 & 2.16 $\pm$ 1.12 & 3.64 $\pm$ 1.67\\
1300 \Msun & 1.07 $\pm$ 0.53 & 2.21 $\pm$ 1.09 & 3.94 $\pm$ 1.89\\
\hline
\end{tabular}
\end{table}

\begin{table}
\caption{SMBH mass and total mass contained in halo MBH averaged over thirty trees.}
\label{tab:total_MBH_mass}
\begin{tabular}{llll} \hline
 \multicolumn{2}{c}{Model} & $\Sigma ~M_{MBH} [M_{\odot}]$ & $M_{SMBH} [M_{\odot}]$ \\
Halo mass & $M_{\bh,seed}$  & & \\
\hline \hline
$1.6\times10^{10}$ & 260 \Msun & $(4.9 \pm 0.9)\times10^4$ & $(1.7 \pm 0.7)\times10^4$\\
 & 1300 \Msun & $(2.0 \pm 0.8)\times10^5$ & $(9.0 \pm 4.1)\times10^4$\\
$1.6\times10^{11}$ & 260 \Msun & $(3.4 \pm 1.5)\times10^5$ & $(1.9 \pm 1.0)\times10^5$\\
 & 1300 \Msun & $(2.5 \pm 0.7)\times10^6$ & $(9.2 \pm 6.2)\times10^5$\\
$1.6\times10^{12}$ & 260 \Msun & $(5.7 \pm 0.8)\times10^6$ & $(1.7 \pm 0.9)\times10^6$\\
 & 1300 \Msun & $(2.3 \pm 0.7)\times10^7$ & $(9.0 \pm 3.5)\times10^6$\\
$1.6\times10^{13}$ & 260 \Msun & $(3.9 \pm 1.2)\times10^7$ & $(1.8 \pm 0.9)\times10^7$\\
 & 1300 \Msun & $(2.6 \pm 0.5)\times10^8$ & $(8.4 \pm
4.6)\times10^7$\\
\hline
\end{tabular}
\end{table}

\begin{figure*}
\hbox{\epsfxsize 8.5cm
\epsfbox{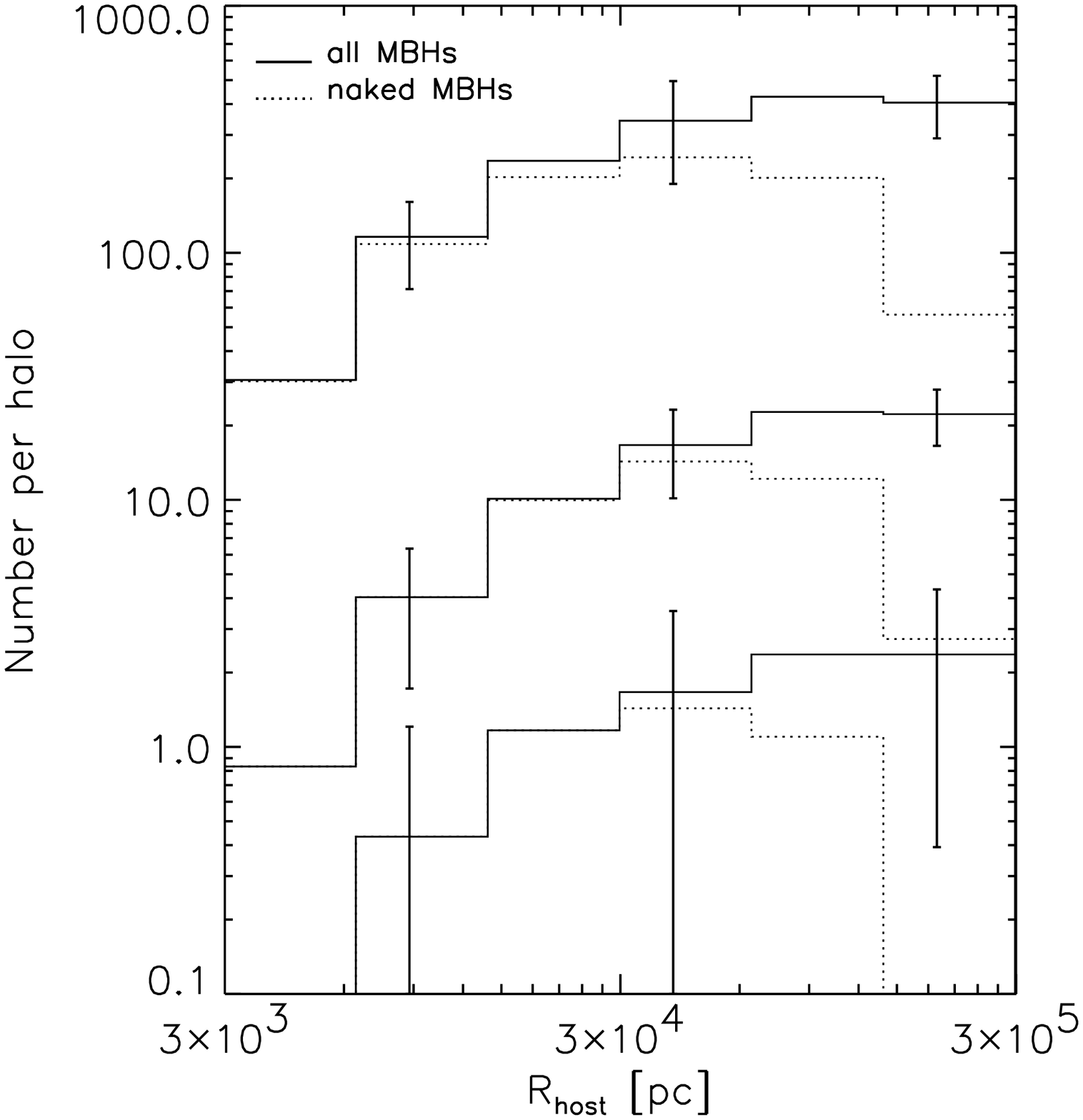}
\hskip 5pt
\epsfxsize 8.5cm
\epsfbox{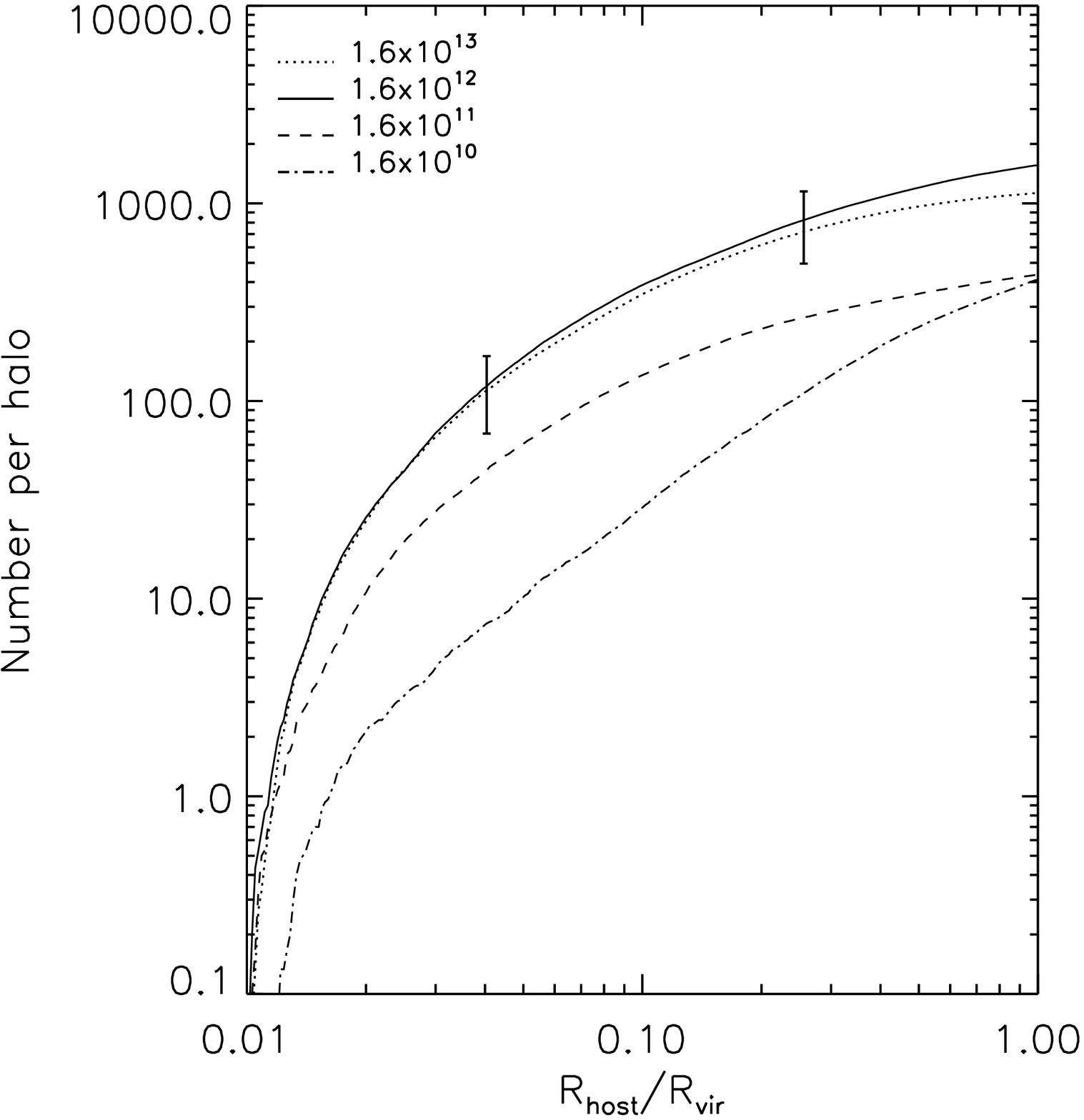}}
\caption{~Radial distribution of MBHs for the case of $260$ \Msun
~seed MBHs. The left panel shows the
differential distributions for all (top set of curves) MBHs and those with masses
above $10^4$ \Msun ~(middle) and $10^5$ \Msun ~(bottom).
The same but only for `naked' MBHs is shown by the dotted lines.
The total number of MBHs within a given distance from the host centre is 
shown on the right for all final halo masses where
distances have been scaled to the virial radius of the respective halo.}
\label{fig:abund_radius}
\end{figure*}

Figure \ref{fig:abund_radius} shows the number of MBHs
as a function of distance from the host centre in the
$1.6\times10^{12}$ \Msun halo. We have only plotted the case
$M_{\bh,seed} = 260$ \Msun~as is essentially the same 
for the two different seed MBH masses.
The left panel indicates that the relative distribution of MBHs with distance
from the halo centre is remarkably similar for the different masses of
MBHs. It also becomes clear that by far most
of the MBHs in the inner part of the halo, are 
naked, that is they have no associated satellite halo.
In the right panel we have plotted the cumulative radial distribution
and we see that it is very similar for the different final halo masses
when scaled to their respective virial radii. Apart from the different 
halo masses that account for different normalisation of the abundance 
of MBHs, the difference in shape, especially for the
$1.6\times10^{10}$ \Msun~halo, likely reflects the higher concentration 
of the halo potential.

In table \ref{tab:local_abund} we have listed the average abundance of 
MBHs in local Earth centred volumes.
Virtually all of these will be seed BHs that have not yet merged and
in the absence of any growth process other than hierarchical merging 
their mass will be equal to that of the initial seed BHs.

The total mass contained in halo MBHs is shown in table
\ref{tab:total_MBH_mass} and compared with the average mass of the
central SMBH. Regardless of seed MBH mass we find that on average the central
SMBH has about 30 to 50 per cent of the mass that is contained in lower mass MBHs in
the galactic halo. 

Within the variance quoted we expect the number and mass abundance of
MBHs particularly in the $1.6\times10^{12}$ \Msun halo as
shown above to be representative for Milky-Way-sized galaxies in
currently favoured $\Lambda$CDM cosmologies.

\subsection{Constraints on initial MBH mass function}
The above results for the two different seed MBH masses give some
indication of the effect of other changes in the masses and numbers of seed MBHs in the
primordial haloes. 

We have seen above that the MBH mass functions are shifted along the
$M_{mbh}$ axis in proportion to the mass of the seed MBHs. This mass,
however, cannot be higher than the total baryonic mass contained in the
original $3 \sigma$ haloes. This translates into the solid grey line shown
in figure \ref{fig:abund_mass} and thus any mass for a single seed MBH 
between 260 and $1.3\times10^4$ \Msun will lead to a present-day MBH
mass function between the upper and lower most ones shown.

By conservation of mass \footnote{Strictly the masses of two merging
BHs are not conserved, but will be lower by a few per cent,
since gravitational waves can radiate away some of
the BHs' rest mass energy. In the following we assume that this effect 
only changes our results by a negligible amount,
although the mass loss through gravitational radiation accumulated in
many mergers for some MBHs may become significant.}, if the  
primordial halo contains more than one MBH of different masses in
the range 260 \Msun $< M_{MBH} <$ 13000 \Msun then the resulting
mass function will again lie between the bottom and the top one shown, but
will have a different slope.
If initially one or more MBHs were present with masses lower than $260 
{\rm M_{\odot}}$, the present-day mass function will correspondingly extend to
lower masses, but will otherwise still be limited by the top mass
function.
This means, that even though we had initially made a fairly specific
choice for the initial MBH mass function in the primordial haloes, any
general form for the MBH IMF is expected to lead to results within the 
limits provided by the mass functions shown, if there is at
least one seed MBH of $260$ \Msun or larger.

We need to stress that the above depends on the  assumption that all MBHs falling to
within one per cent of the virial radius merge efficiently in all haloes 
merging along the way to produce the final host halo. We consider the
implications of less efficient or no merging of MBHs in the next
section.

If the only or at least most significant source of seed MBHs is that
forming in the $3 \sigma$ haloes then the total mass contained in halo 
MBHs can be used to normalise 
the initial mass function of seed MBHs, to which it is related by the
background cosmology. The latter determines the average merger history
of haloes and thus the average number of $3 \sigma$ haloes ending up in more massive 
haloes later on. Note that this is not much affected by the merger efficiency
of MBHs since the present-day MBH mass function is dominated by seed
MBHs that have not merged, and that contribute a similar amount to the total mass contained
in halo MBHs as the few very massive MBHs that have resulted from
multiple mergers of seed MBHs.

\subsection{MBH merger efficiency}
Up to now we have considered any MBH as having merged with the central 
MBH, when it comes within  one per cent (hereafter referred to as the
merger region) of the virial radius of the
host halo at that time.
There are various ways in which the actual merger efficiency could be
lower than this, and so our results above only provide an upper limit
on how much the MBH merger process can contribute to the mass of
central and halo MBHs.

One major source of inefficiency is of course the time it takes for
any MBH to spiral into another and typically more massive  MBH at the
centre of their common host and how likely it is then for the two to
merge. One does not necessarily imply the other -- at early
times haloes are smaller, that is at the first encounter 
the two central MBHs within any two haloes will start out much
closer and so are more likely to spiral to the common centre of the
halo merger remnant in a relatively short time.
Because there are more low mass haloes this might then give rise to configurations consisting of 
more than two MBHs and thus the possibility of sling-shot
ejections. In other words some fraction of MBHs, although having
travelled to the centre quickly might eventually end up being expelled 
rather then merging. 
 This has implications for the most
massive trees. Haloes at the resolution limit in these trees have a
mass above $M_{min}$ and therefore might appear in the tree with
several seed MBHs which we have thus far assumed have merged to form
one MBH (c.f. section 3.2). This may no longer be the case if
slingshot ejections occur. Assuming that in this case the lightest
MBHs are ejected, however, this should not significantly reduce the
mass of the central MBH.

\begin{figure}{}
      \centerline{\resizebox{\hsize}{!}{\includegraphics{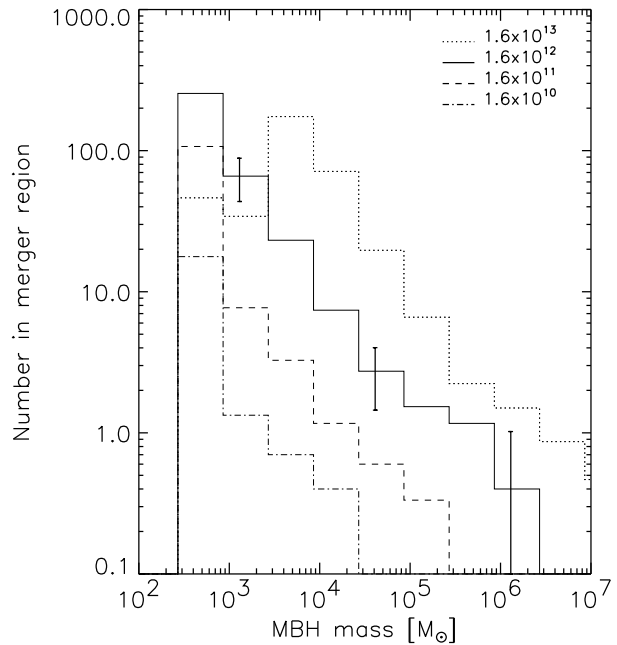}}}
      \caption{Mass function of MBHs that have fallen into the merger region for all final 
halo masses and $M_{\bh,seed} = 260$ \Msun.}
      \label{fig:mergregabund}
\end{figure}

We can also ask what happens if those MBHs that have crossed 
into the merger region of the host do not actually merge at all but keep orbiting on only mildly radial orbits
with associated long orbital decay time scales. We will subsequently refer to this as the `no-merger' scenario.
The first and most crucial consequence of this
is that a SMBH grown through hierarchical mergers of these MBHs 
would not exist in the first place.  Instead the SMBH mass would simply add to the
total mass contained in halo MBHs.
In figure \ref{fig:mergregabund} we show the mass function of MBHs
that have fallen into 
the merger region, and which in the no-merger scenario would just remain
orbiting there. Their total number is about a quarter that of the
MBHs in the haloes as given in table \ref{tab:global_abund}.

\section{Detections} \label{sec:mbhdetect}

\subsection{X-rays}
An abundance of massive black holes as determined above should be
detectable in various ways. 
First and foremost we expect these MBHs to be sources of
X-rays. These could arise as a result of accretion from the interstellar
medium as it moves through the host halo \cite{fujita98}. This effect is only expected to be
large for MBHs travelling through the disk or bulge at relatively low
speeds. However, above we have seen that by far most MBHs are actually
in the halo and not in either bulge or disk. In this case the
number of significant MBH X-ray sources is therefore expected to be
rather low. 

\begin{figure*}
\hbox{\epsfxsize 8.5cm
\epsfbox{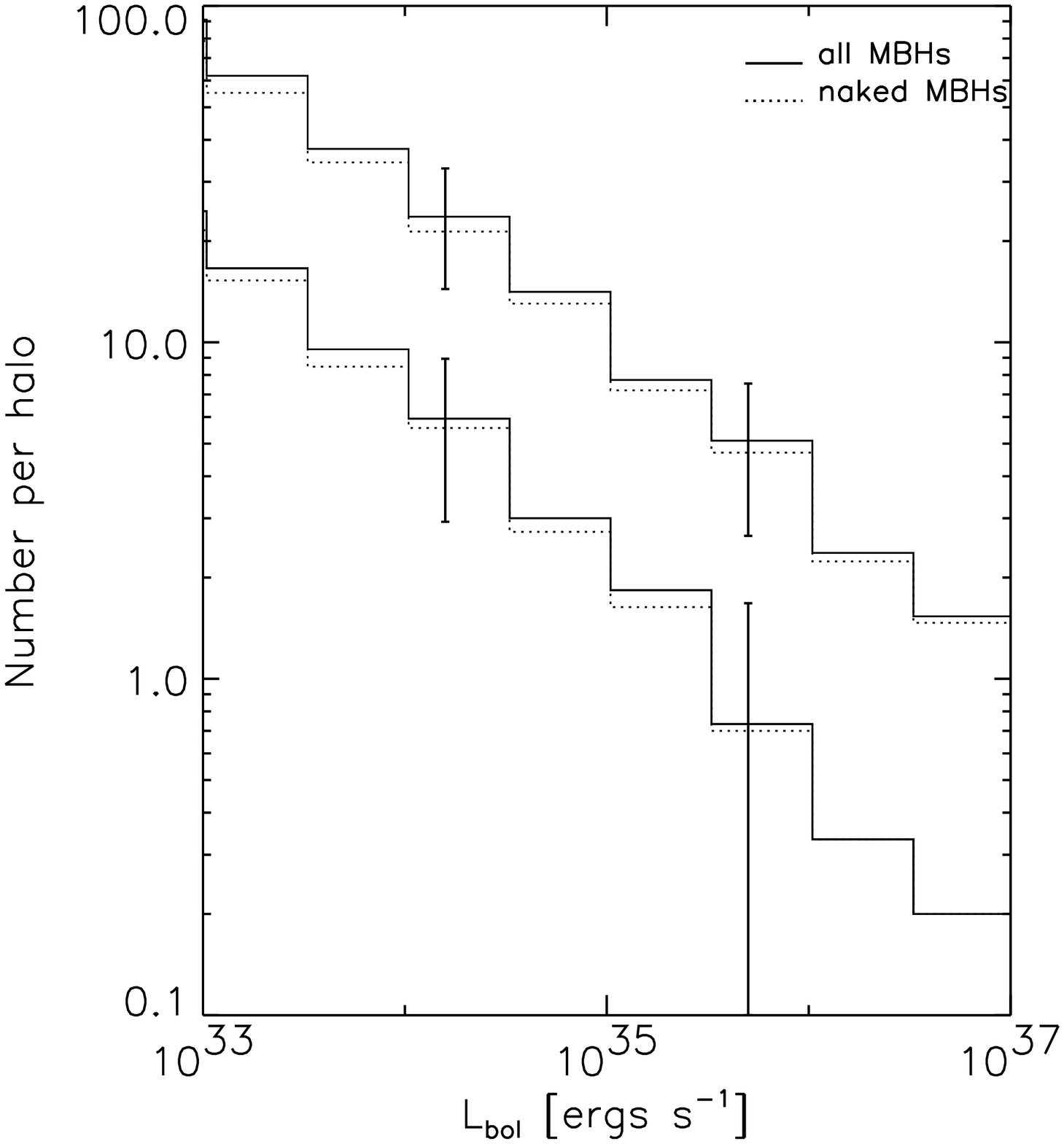}
\hskip 5pt
\epsfxsize 8.5cm
\epsfbox{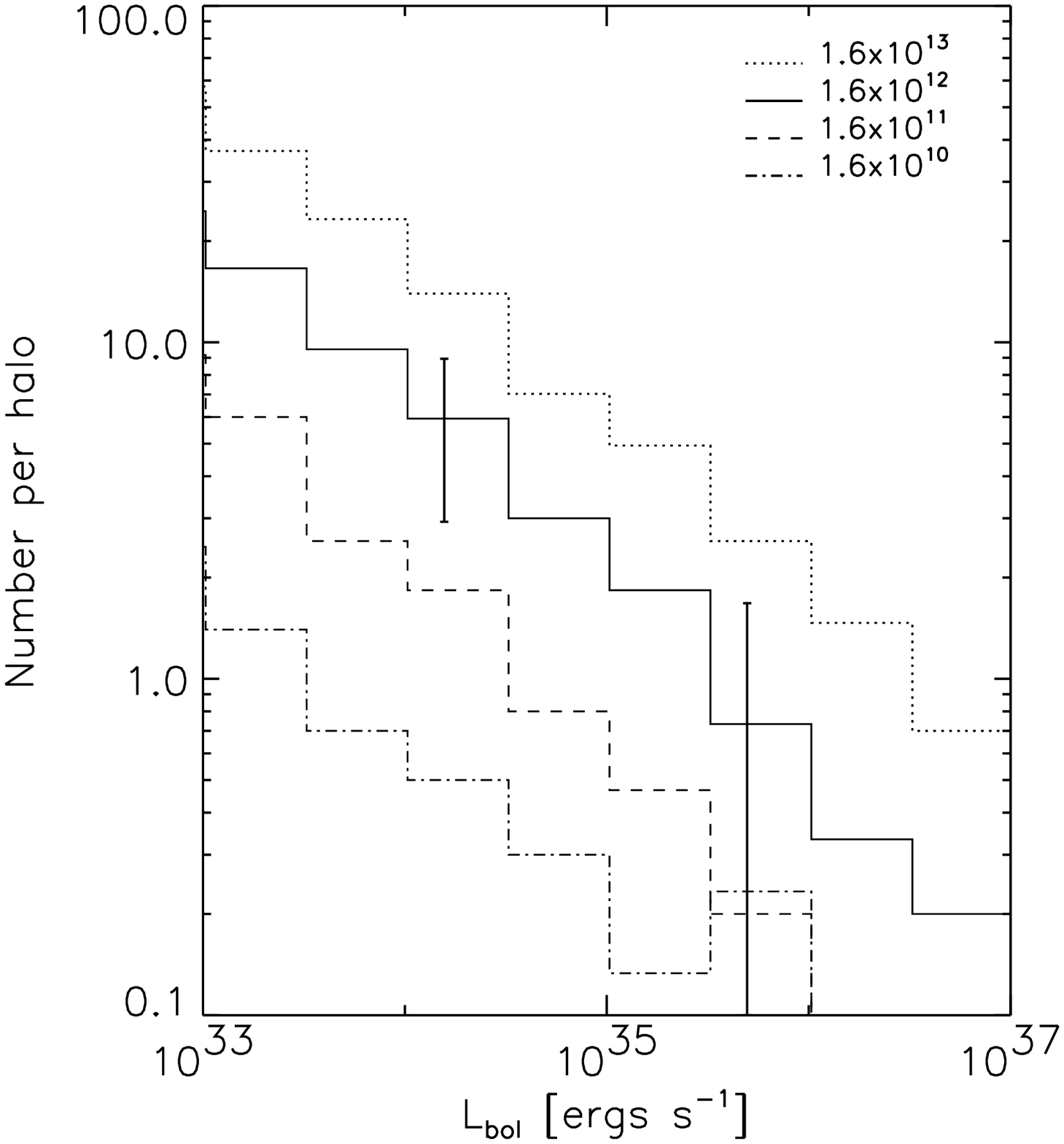}}
\caption{~Bolometric luminosity function for MBHs accreting from the
halo ISM. Results are shown for the $1.6\times10^{12}$ \Msun halo and
all as well as only naked MBHs. Upper and
lower data sets are for $M{\bh,seed} = 260$ and $1300$
\Msun respectively. In the left panel the luminosity functions for
$M{\bh,seed} = 260$ \Msun and all final halo masses are shown. In all
cases the radiative efficiency is $\eta = 0.01$.}
\label{fig:ISMXlum}
\end{figure*}

We have estimated this using the Bondi-Hoyle
\cite{bondi44,bondi52} mass accretion rate
and the standard radiative efficiency for thin disk accretion, $\eta = 0.01$. The resulting
bolometric luminosity function is shown in figure
\ref{fig:ISMXlum}. Depending on the accretion model (e.g. advection or convection
dominated accretion flows, ADAF or CDAF \cite{manmoto97,ball01}) X-rays will
account for 5 to 30 per cent of this luminosity.
Most of the very luminous sources are `naked' MBHs, i.e. in tidally
stripped satellites, which implies that they must be orbiting at relatively
small distances from the host centre and therefore in or close to the
bulge region.

\begin{figure*}
\hbox{\epsfxsize 8.5cm
\epsfbox{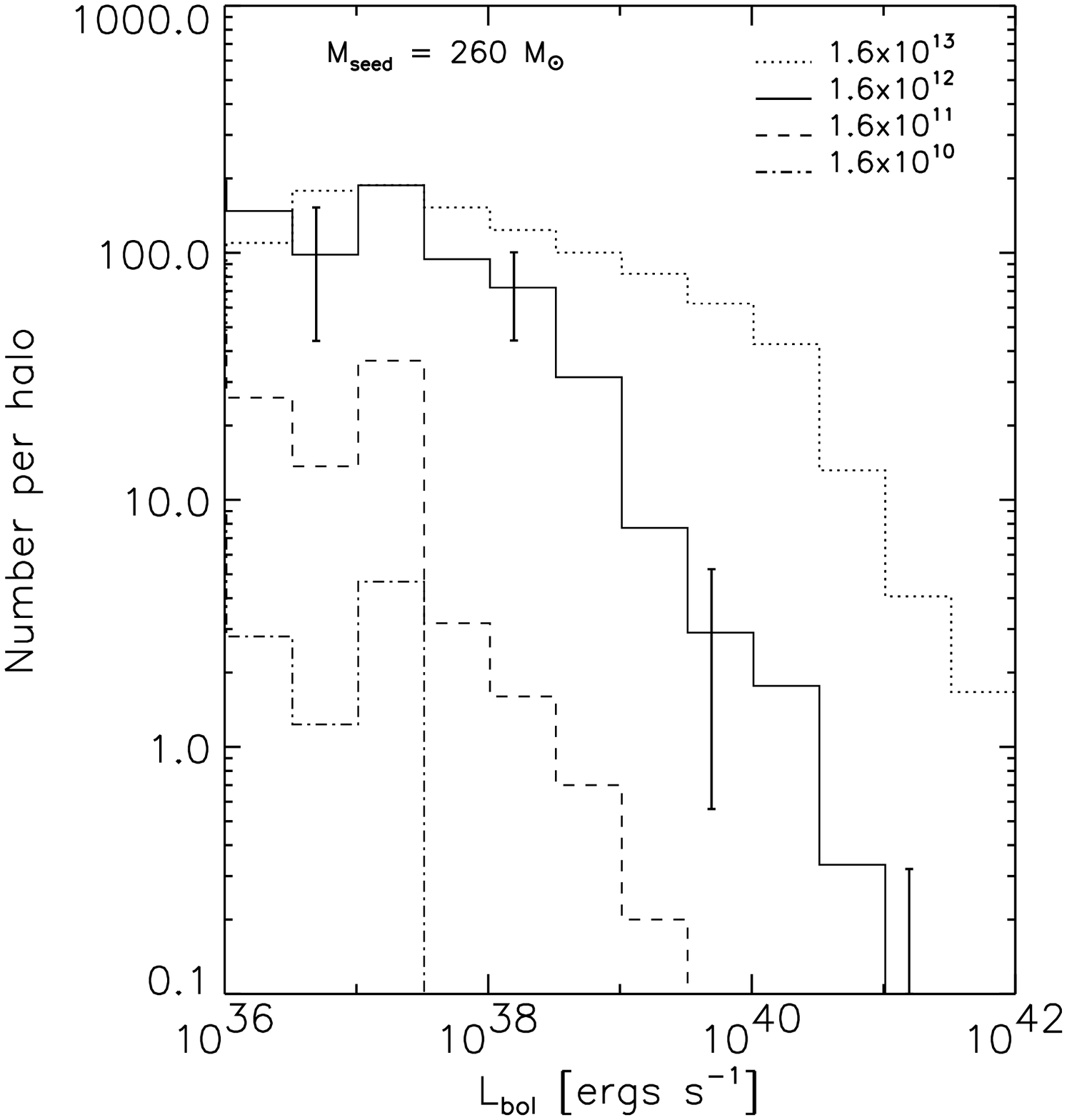}
\hskip 5pt
\epsfxsize 8.5cm
\epsfbox{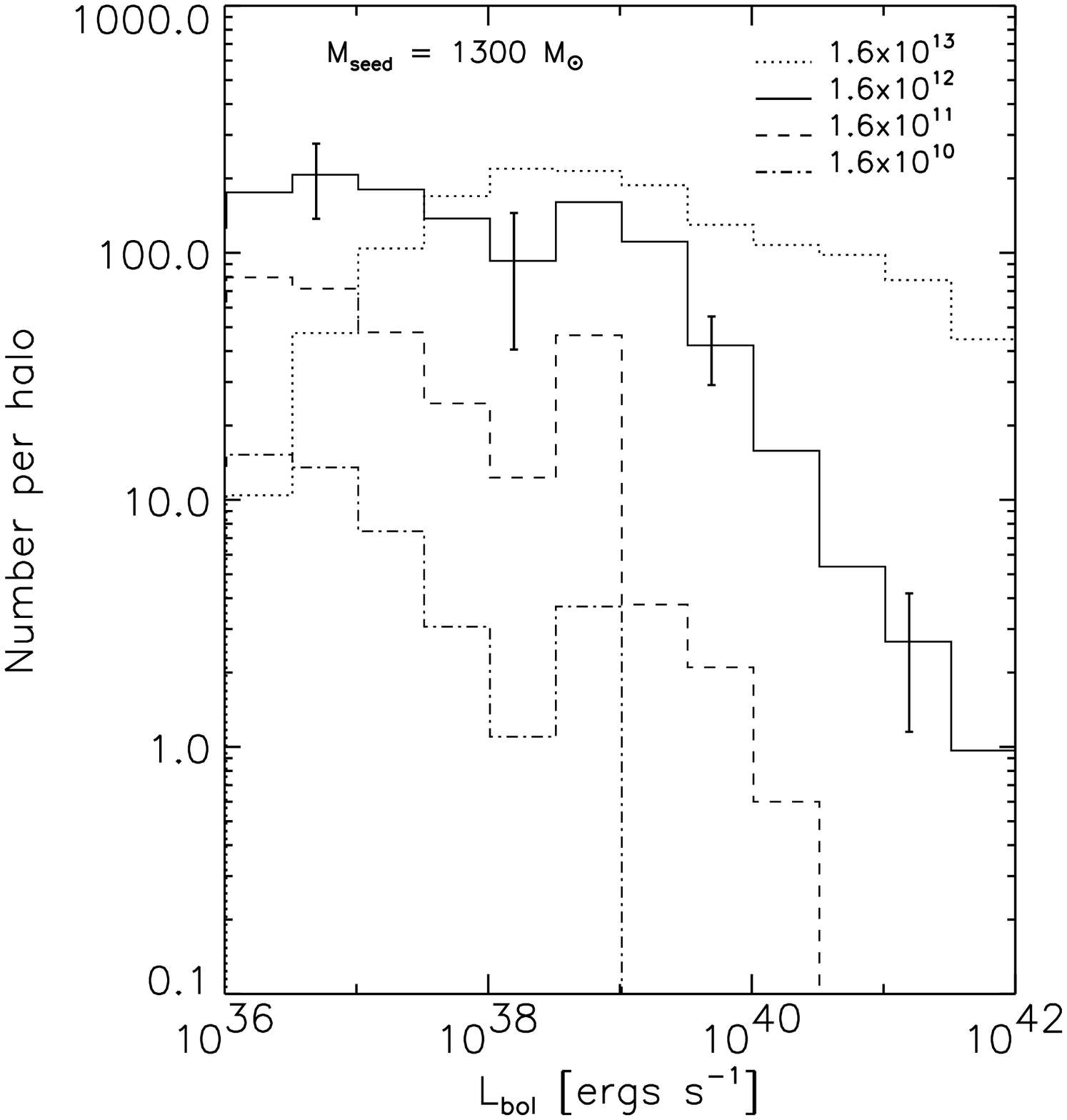}}
\caption{~Bolometric luminosity function for MBHs accreting from
baryonic core remnants of their original satellites with radiative
efficiency $\eta = 0.001$. Results are shown for all final halo masses 
and the two seed MBH masses considered.}
\label{fig:barcorXlum}
\end{figure*}

Another more likely scenario is that most MBHs
will actually remain embedded in a dense baryonic core remnant of their
original satellite halo. If this core
had formed within a satellite before it entered the host, then it is
likely to survive in the host gravitational tidal field even though
most of satellite's dark matter may be tidally stripped.
Our calculations indicate that with this `portable fuel supply' travelling
along with the MBH, accretion rates can be much higher. In particular
we have assumed that the core is a constant density sphere with a radius
that is about 10 per cent the virial radius of the
original satellite and contains all its baryonic mass. Applying the
Bondi-Hoyle accretion formula to this with a smaller radiative
efficiency of $\eta = 0.001$ we obtain much larger luminosities as is
shown in figure \ref{fig:barcorXlum} for the two different seed MBH
masses considered.

\begin{figure}{}
      \centerline{\resizebox{\hsize}{!}{\includegraphics{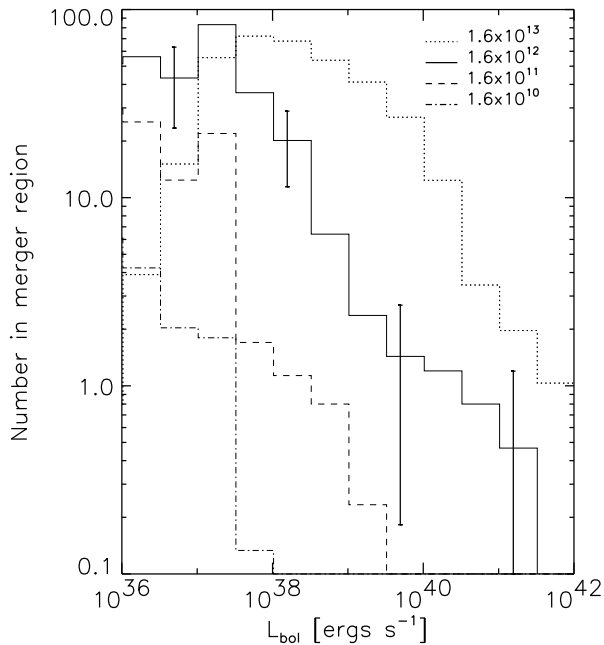}}}
      \caption{Same as figure \ref{fig:barcorXlum} but for MBHs in
merger region and considering they do not merge. Data shown are for
seed MBH mass $260$ \Msun and all final halo masses.}
      \label{fig:mergregXlum}
\end{figure}

Depending on the accretion model this implies X-ray luminosities of up 
to $\sim 10^{41} {\rm erg ~s^{-1}}$, and thus
above the luminosities typical for stellar mass X-ray binary systems.
These should be clearly detectable, since most X-ray emitting MBHs are in the halo
where they remain relatively unobscured by gas and in addition much of
the X-ray luminosity is expected to be emitted in the hard X-ray band
which is even less subject to absorption. 
It seems plausible that the brightest of these could account for the ultra luminous
off-centre X-ray sources observed in some galaxies
\cite{zezas02,kaaret01} as well as an alternative to point sources
interpreted as low luminosity AGN. 
For the no-merger scenario  we have also shown the corresponding
luminosity function in figure \ref{fig:mergregXlum} for the MBHs
orbiting in the merger region 

Since only a small baryonic core is required for X-ray emission we
would not necessarily expect the sources to be embedded in a stellar
cluster or dwarf galaxy. These may have been stripped away or
depending on their star formation history not been present to an
observable level in the first place.  

Depending on the accretion model adopted  MBHs also
emit in the optical / infrared
with a spectrum and luminosity that varies significantly. Because of
this uncertainty and additional problems, such as absorption, optical
or infrared observations of MBHs are probably only useful as a follow
up to X-ray detections.

\subsection{Other detections}
Other types of detection that could in principle be used to probe the
presence and abundance of MBHs are possibly micro-lensing and gravitational
waves.
For Milky-Way sized haloes the abundance of MBHs is too low  (c.f. table 
\ref{tab:local_abund}) and their masses too large to yield a
significant micro-lensing signal over a reasonably short period of time 
at least in current micro-lensing surveys, like the MACHO project. Future
astrometric missions, like GAIA, might be in a better position to
detect MBHs in the solar neighbourhood.
For the no-merger scenario, this  might in fact be the only way to
detect of the order of 100 -- 1000  MBHs in the central kpc.
\\
Detection by gravitational waves, in comparison, is more
straightforward. Detection rates for hierarchically merging central SMBHs
could be calculated out to redshifts larger than 100, if MBHs did exist there, 
\cite{haehnelt94,menou01} and depend sensitively on the merger history 
and abundance and distribution  of seed black holes. In principle MBHs falling into the centre and
merging with the central SMBH will produce gravitational wave
events in addition to those arising from mergers between central SMBHs 
in the wake of major halo mergers. 
In fact, in our model the number of MBH -- central SMBH mergers is
expected to be higher than this, since between any two galaxy mergers
with corresponding mergers of their central SMBHs, there is a number
of MBHs that fall to the centre and coalesce with the central SMBH.
The latter have a lower gravitational wave amplitude
because of the very different masses of MBH and SMBH and detection of these events is
therefore limited to lower redshifts, but should still cover the range 
up to $z \sim 20$, which is where the haloes of the seed MBHs in our model 
would undergo their first mergers..\\  

\section{Summary and Discussion} \label{sec:summary}
We have used a semi-analytical approach to track the merger
history of massive black holes and their associated dark matter haloes,
as well as the subsequent dynamical evolution of the MBHs within the
new merged halo.
In particular we have looked at the possibility that MBHs that are the 
remnants of massive population III stars, forming in low mass haloes at 
redshifts $z \sim 24$, could hierarchically build up to contribute to
the present-day abundance of central galactic SMBHs.
If this is the case then a number of remnant MBHs is expected to orbit 
inside galactic haloes.
Although our analysis has been carried out for one of the currently
favoured $\Lambda$CDM cosmological models, we expect our findings to
hold for any model that provides for hierarchical structure formation
such as CDM models in general, but notably excluding Warm Dark Matter
and other models with a cut-off or discontinuity at some
specific scale in their corresponding cosmological  matter power spectrum.
\\
The main findings of our analysis are:
\begin{enumerate}
\item For Milky-Way sized galaxies, of the order $10^3$ MBHs that have
not reached the host centre are expected to orbit within the
halo. Around 1/3 of these will be seed mass MBHs, 85 per cent of these MBHs
with masses up to $10 \times M_{\bh,seed}$.

\item For a seed MBH mass of $260$ $(1300)$ \Msun ~some 5 to 8 (2 to 3) MBHs with masses 
above $10^5$ $(10^6)$ \Msun ~are expected in the halo of Milky-Way sized galaxies. 

\item Hierarchical merging of seed MBHs with masses 
of $M_{\bh} \sim 10^3$ \Msun forming in haloes collapsing
from $3 \sigma$ peaks in the matter density field at
$z \sim 24$ can contribute up to 10 per cent to the present-day mass
density contained in SMBH. Another mechanism for the SMBH to gain mass,
such as gas accretion, appears inevitable. 

\item Depending on the size of a baryonic core remnant around
the MBHs, they could be significant sources of X-rays and possibly account for
the ultra-luminous off-centre X-ray sources that have been
found in a number of galaxies. Accretion from the host ISM is probably 
not important.
\end{enumerate}
We find that the mass functions
for all seed MBH masses considered are essentially the same and only
shifted along the mass axis proportional to the mass of the seed MBHs.
This is because it is the mass of the satellite haloes and not that of the MBHs that
dominates their dynamical evolution in a host halo.

Our findings are consistent with the results of another recent
investigation by Volonteri, Haardt \& Madau (2002). We find the
total mass density in a Milky-Way sized galactic halo is about a
factor 10 higher than their value inferred from the density function
of `wandering' BHs in galactic haloes. This is what we would expect on the basis of the
difference in seed BH masses and height of the peaks where initial
collapse occurred (their $3.5 \sigma$ vs our $3 \sigma$).

Furthermore, we find the mass of a central SMBH in a Milky-Way sized
halo to be $1.7 \times 10^6$ \Msun. Accounting for the difference in
seed MBH masses used this agrees to within a factor 2
with the central SMBH mass of $\sim 5\times 10^5$ \Msun for a galaxy
sized halo ($\sigma \sim 100 {\rm ~kms}^{-1}$) as implied by
their $M_{\bh} -\sigma$ relation (with no gas accretion).
This also happens to coincide with the mass determined for the SMBH in
the Milky Way, although the Milky Way SMBH is known to lie significantly below the
observed $M_{\BH} - \sigma $ relation.

However, the slightly non-linear $M_{\BH} - M_{bulge}$ correlation corresponds to a
$M_{\BH} - \sigma $ relation whose logarithmic slope ($\sim 4.0$) does
not match the much flatter one they determined ($\sim 2.9$) for
$3 \sigma$ collapse and no gas accretion. 
We believe this to be primarily a result of the different
assumptions made about the MBH merger process. In particular the
inclusion of triple BH interactions and sling-shot ejections, that
they find, would probably lead to even lower central SMBH masses in
our analysis.

While the fiducial model of Volonteri ~\ea is based on the collapse of
$3.5 \sigma$ peaks and a seed MBH mass of 150 \Msun, we consider $3 \sigma$ peaks and a higher mass for 
the seed MBHs, both of which imply a higher mass density in MBHs at
high redshift. This in turn means that less gas accretion onto MBHs is needed to match the
$M_{\BH} - M_{bulge}$ that is actually observed in nearby galaxies. In 
any case even for our lightest seed MBH mass considered, any resulting 
central galactic SMBH in our analysis would need to accrete at least
50-100 times its own initial mass to match the observed relation. This 
is in accord with a number of studies (see e.g. Yu \& Tremaine 2002
and references therein) that find the present day SMBH mass density to
be consistent with the amount of gas accreted during the optically bright
QSO phase. 
On the other hand gas accretion during the QSO phase alone cannot
explain growth from a stellar mass BHs to the most massive SMBHs ($\ge 10^9$ \Msun). Even  
if stellar mass BHs are accreting at he Eddington limit, the QSO phase
would not last long enough to accommodate the required number of e-folding times for the
BHs to grow to SMBH size. The need for intermediate mass seed BHs
and/or some merging of MBHs/SMBHs is therefore necessary to explain
the presence of the most massive SMBHs.

Our numerical results depend on a number of parameters that are not
yet well constrained, notably the exact height of the fluctuations in
the matter density field that are supposed to collapse to form the
first baryonic objects and the initial mass function of metal poor
stars forming inside these. While the former could possibly be
determined better by improved numerical simulations, we have shown
that, particularly for the abundance of MBHs in the halo, our results
hold qualitatively for a wide range of different  IMFs.

If the halo MBHs could be uniquely identified by their X-ray emission
or otherwise, then within the context of our model they could also be used
to tag (remnants of) substructure orbiting in a galactic halo. In this 
way they would complement counts and location  of dwarfs and star
clusters as measures of substructure in the galaxy and the halo.

Our results for the growth and present-day mass of the central SMBHs do 
depend sensitively on how efficiently MBHs merge at the host
centre. Here we have taken the view that during major mergers any MBHs 
orbiting within the core region of the host will be dragged towards the 
central SMBH quickly, aided by the massive inflow of gas. Due to the
increased non-homogeneity, violent dynamical evolution and departure
from spherical symmetry during this phase, analytical estimates of dynamical time scales
presumably overestimate the time required for MBHs to travel to the centre. 

However, a more detailed analysis of this process will be required for 
the calculation of event rates of mergers between central and
inspiralling MBHs and the accompanying gravitational wave emission.
\vspace{0.5cm}
\\
The authors wish to thank R. Bandyopadhyay and G. Bryan for useful
discussions and are grateful to the referee for a number of helpful comments.
RRI acknowledges support from Oxford University and St Cross College, Oxford.
JET acknowledges support from the Leverhulme Trust.

\end{document}